\documentstyle[epsfig,prb,aps,eqsecnum]{revtex}

\begin{document}

\draft
\twocolumn[\hsize\textwidth\columnwidth\hsize
\csname @twocolumnfalse\endcsname
\title{
{\vspace{-7mm}\small\framebox[14cm][l]{%
Submitted to Phys.Rev.Lett., 24.3.1998, resubmission 26.5.1998} \\
\vspace{2mm}}%
Carrier doping in Pr$_{1+x}$Ba$_{2-x}$Cu$_3$O$_7$ studied by NMR}
\author{M.W. Pieper, F. Wiekhorst}
\address{Universit\"at Hamburg, IAP, Jungiusstra\ss e 11, D-20355 Hamburg,
e-mail: Pieper@Physnet.Uni-Hamburg.de}
\author{T. Wolf}
\address{Forschungszentrum Karlsruhe, Institut f. Technische Physik,
D-76021 Karlsruhe, Germany}
\date{\today}
\maketitle
\begin{abstract}
We investigated the type of the carriers doped in the CuO$_2$-planes of
non-superconducting Pr$_{1+x}$Ba$_{2-x}$Cu$_3$O$_7$ single crystals
by nuclear magnetic resonance of Pr and chain-site Cu. The
different spectra in the solid solution system show that the holes in
Pr-rich crystals are localized in states similar to the ones in
stoichiometric samples, that is in the $4f^2-2p\pi$-hybridization band
proposed by Fehrenbacher and Rice. In contrast, the spectra of crystals
prepared under conditions favoring Ba-rich phases indicate the presence
of holes in Zhang-Rice singlets in the CuO$_2$-plane. Our results
strongly support the model that superconductivity reported recently for
Pr123-crystals is due to Ba-rich regions in the inhomogeneous samples.
\end{abstract}
]

\narrowtext

PrBa$_2$Cu$_3$O$_7$ has been investigated for more than a decade because
it is the only antiferromagnetic insulator in the isostructural
($R123$-)family of superconducting cuprates ($T_c=92\pm 2$~K for
$R$Ba$_2$Cu$_3$O$_7$, $R$=Y, La, rare earths). The most common model for
this exceptional property of Pr123 assumes a carrier localization in
antibonding oxygen orbitals hybridized with the $4f^2$-shell of Pr, the
so-called Fehrenbacher-Rice (FR) states\cite{feh198,lie197}. As a
consequence the concentration of Zhang-Rice (ZR) singlet hole carriers
in the CuO$_2$-planes is reduced below the threshhold of
superconductivity\cite{maz357,mer358}. Numerous experimental
investigations since 1987 indicated that superconductivity may be
restored in Pr123 only by dilution of the Pr-sublattice\cite{rad216}.
Isovalent substitution, e.g. by Y, leads to an (AF) insulator- (SC)
metal transition above app. 40\% Y \cite{Zitat1,Zitat2}. As expected,
this dilution effect is enhanced by additional hole doping if one
substitutes Pr$^{3+}$ by Ca$^{2+}$.

This situation seemed to change when Zou et al. prepared pure
superconducting Pr123-crystals with $T_c$-values above 80K ($> 100$~K
under pressure)\cite{zou328}. They considered two reasons for the
occurence of superconductivity in their crystals, both due to the
special preparation technique. First, the hybridization could be
suppressed in a stoichiometric modification of Pr123, as indicated by
the longer Pr-O(2)-bond length. This is, however, very difficult to
verify in the inhomogeneous crystals. The second possibility is a
Ba-substitution of Pr, favouring superconductivity in a similar way like
Ca does. The occurence of nonstoichiometric phases
R$_{1+x}$Ba$_{2-x}$Cu$_3$O$_7$ especially in the case of light rare
earths is known since 1987, when Takekawa et al. reported the
substitution of Ba by the rare earths with large radii\cite{Zitat3,Zitat4}.
Before the result of Zou et al. it has been assumed generally that $x$
is always positive. If superconductivity in Pr123 is in fact due to
$x<0$ one expects to observe holes in the FR states in Pr-rich samples,
and in ZR singlets in Ba-rich crystals. In order to verify this
expectation we studied the nuclear magnetic resonance (NMR) of Cu(1)-
and Pr-nuclei in Pr$_{1+x}$Ba$_{2-x}$Cu$_3$O$_7$. The magnetic field
($B_{hf}$) and the electric field gradient (EFG) tensor ($V_{ij},
i,j=x,y,z$) of these sites allow not only to distinguish between a
doping of the chain- or plane-layers\cite{gre354,inpreparation}, but
also between a doping of the planes by ZR singlets, which may lead to
superconductivity, or by FR states, which stay localized. The
dependence of the spectra on the preparation conditions should give,
therefore, insight into the origin of superconductivity in Pr123.

The so-called solid solution ($R123ss$) or homogeneity range in
principle is a four-dimensional volume in the phase diagram spanned by
temperature $T$, partial oxygen pressure $p(O_2)$, and the (2D-)
triangle of cation concentrations sketched for fixed $T$ and $p(O_2)$ in
fig.1. The allowed, stable compositions form the surface of this volume
with the ellipses in fig.1 indicating cross sections. In the absence of
other metals which may substitute for Cu (e.g. Al from the crucibles)
one may assume a constant Cu content. In this case the ellipses in fig.1
degenerate to horizontal lines at the concentrations
R$_{1+x}$Ba$_{2-x}$Cu$_3$O$_7$, and the homogeneity range has only the
three degrees of freedom $T$, $p(O_2)$, and $x$.

The accessible range of $x$ depends on temperature, oxygen partial
pressure, and on the rare earth radius. In general, $x$ increases with
increasing oxygen partial pressure and rare earth radius, and with
decreasing temperature, at least in the vicinity below the peritectic
temperature of the $R123ss$. Accordingly, the largest $x$ in air
atmosphere was observed in the La123ss system\cite{Zitat5} ($x=0.7$),
whereas $x$ is below 0.01 for rare earth ions smaller than Gd.

We grew Pr123ss single crystals with different Pr/Ba ratios from CuO/BaO
fluxes in MgO crucibles by the slow cooling method\cite{Zitat6}. The
Pr/Ba ratio of the growing crystals was set by the Cu/Ba ratio of the
flux via the conodal line to the solidus surface. Temperature and oxygen
pressure were chosen to obtain non-superconducting crystals with $x>0$,
$x=0$, and, if possible, $x<0$ (see tab.1), denoted below Pr$^+$123,
Pr$^0$123, and Pr$^-$123,
\begin{figure}
\psfig{file=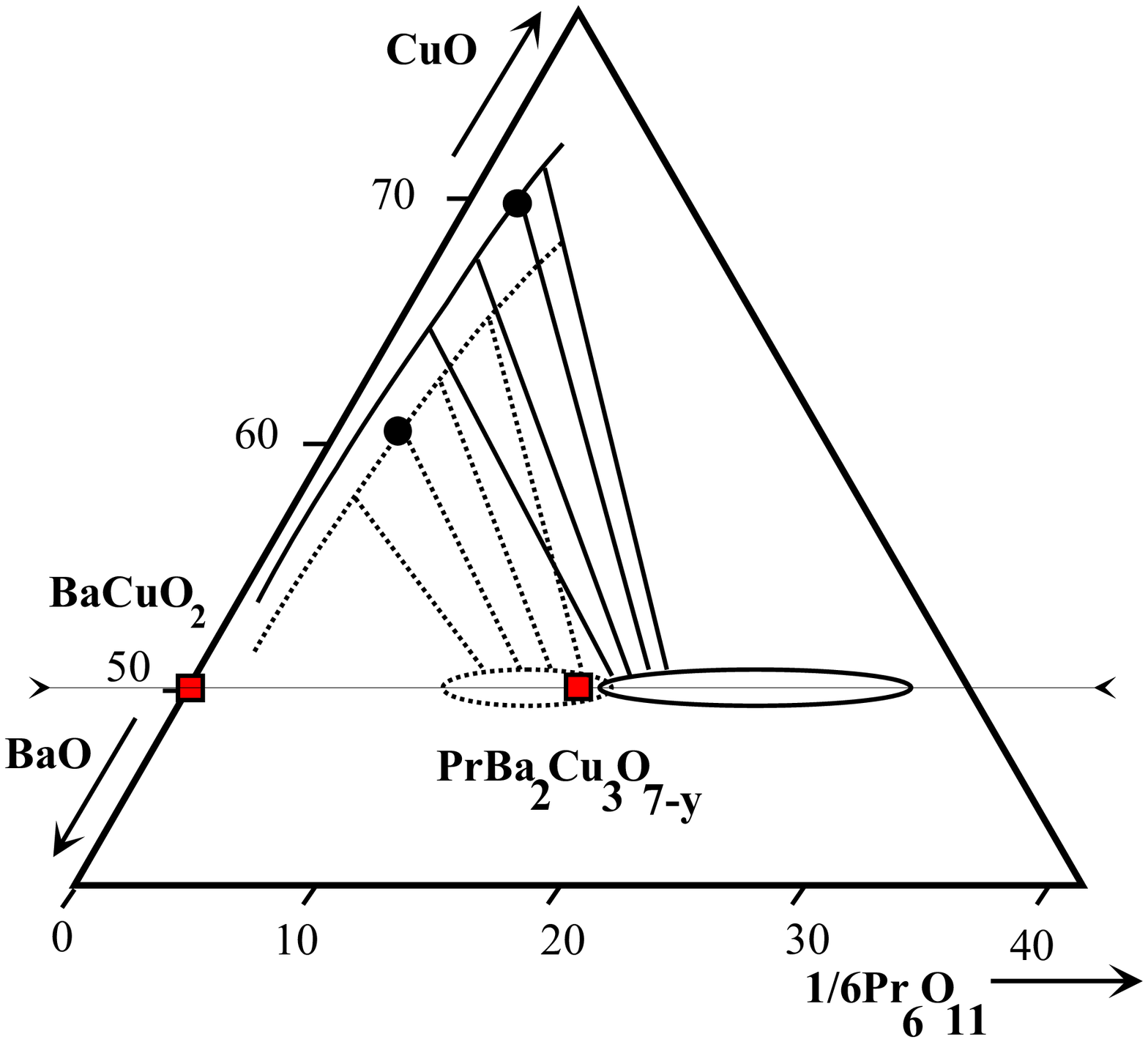, width=8cm, clip=}
\psfig{file=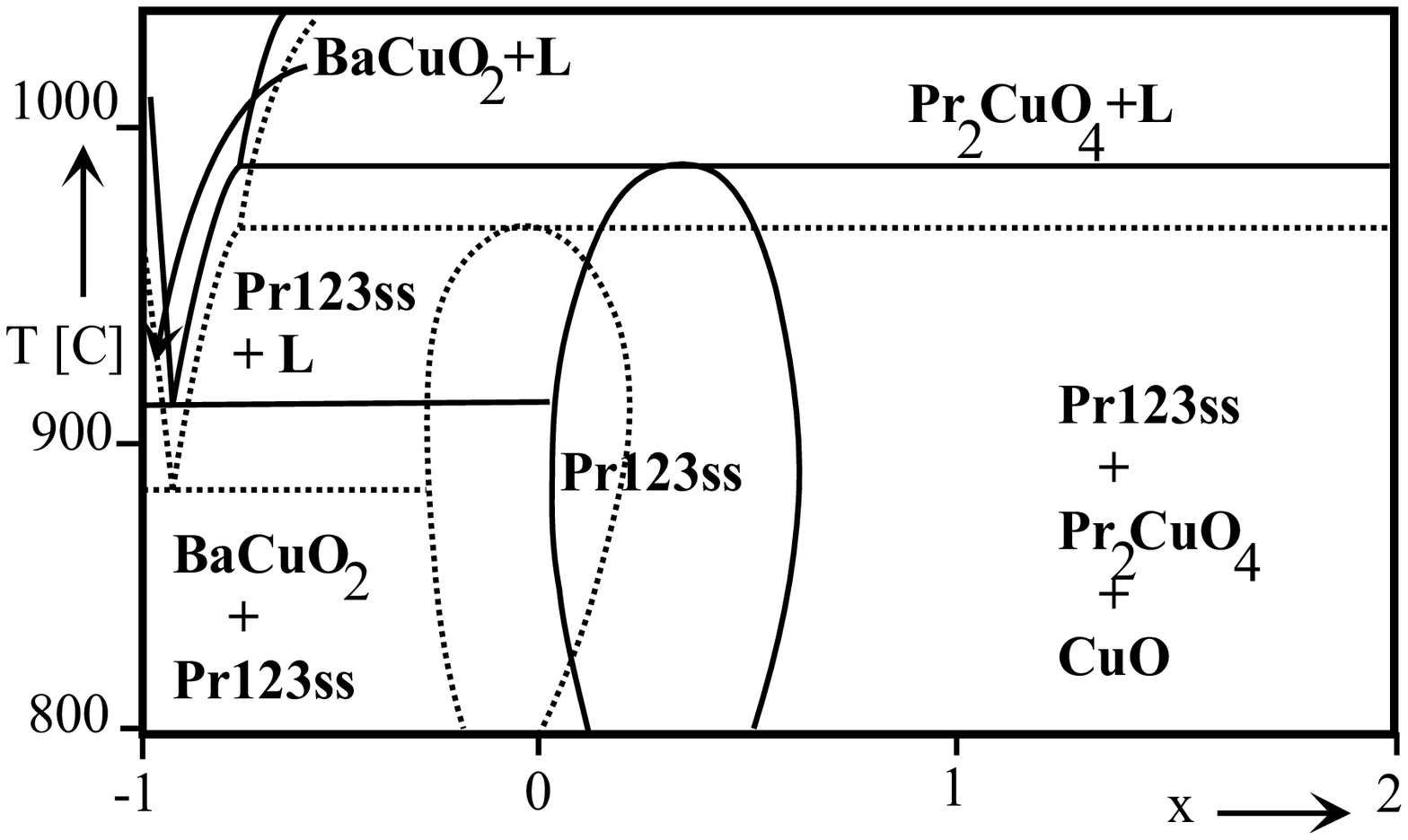, width=8cm, clip=}
\caption{Top: Schematic phase diagram of the Pr-Ba-Cu-O system at
940$^\circ$C in 1 bar O$_2$ (solid lines) and in 63 mbar O$_2$ (dotted
lines). Stoichiometric phases are indicated by filled squares. A few
conodal lines connecting the liquidus surface with the $R$123ss are
shown with the starting compositions for the crystal growth marked by
the dots. Bottom: Schematic phase diagram along the horizontal line
corresponding to Pr$_{1+x}$Ba$_{2-x}$Cu$_3$O$_{7-y}$ in the upper part,
again for 1 bar O$_2$ (solid), and 63 mbar O$_2$ (or 300 mbar air,
dotted).}
\label{Pr123ss}
\end{figure}
\noindent
respectively. We note that we expect $x\approx
0.05$ from neutron work on crystals from the batch of
Pr$^0$123.\cite{mar336} After growth the remaining flux has been
removed with a porous piece of ZrO$_2$ ceramic. The crystals were cooled
to room temperature and, finally, oxidized in 1 bar flowing oxygen in
the temperature range 600--300$^\circ$C. Neither
susceptibility measurements down to 6~K nor the NMR experiments to 1.3~K
described below gave any indications for superconductivity in our
crystals.

\begin{figure}
\psfig{file=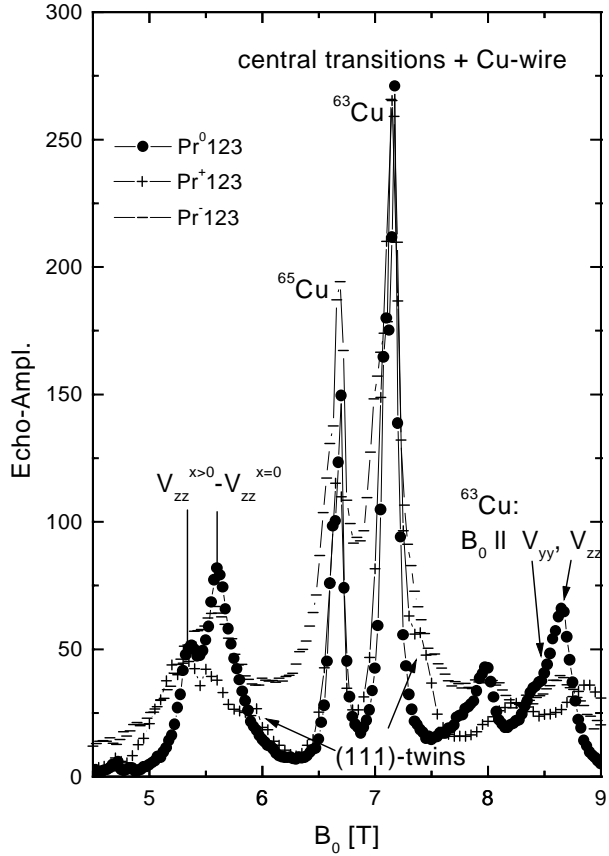, width=8cm, clip=}
\caption{Cu(1)-field-sweep spectra in Pr$_{1+x}$Ba$_{2-x}$Cu$_3$O$_7$ at
81~MHz, 4.2~K with $B_0\| a,b$. See tab.1 for the preparation
conditions. 
At the high-field satellites of Pr$^0$123 the assignment of the
$^{63}$Cu-lines to the orientation of the twins in the field is
indicated, showing that $\mid V_{zz}\mid\approx\mid V_{yy}\mid$. At the
low field satellites the difference in the splitting between Pr$^+$123
and Pr$^0$123 is marked. The two shoulders assigned to (111)-twins appear
at the positions expected if the field is along the space diagonal of
the EFG-tensor, supporting the existence of these twins in the crystal.
}
\label{CuRes}
\end{figure}

The symmetry and the morphology of the oxidized crystals depend in a
characteristic way on $x$ that we also find in the La123ss and Nd123ss.
Pr$^-$123 and  Pr$^0$123 exhibited an orthorhombic structure with twins,
whereas Pr$^+$123 kept its tetragonal structure even after additional
oxidation treatments under high oxygen pressures. This result was
expected because Pr ions on the Ba site force extra oxygen ions on the
O(5) sites between the Cu(1)O-chains,\cite{mal000} which reduces the orthorhombicity.
The morphology of the crystals changes with increasing $R$/Ba-ratio from
isometric blocks, sometimes with additional (101)-faces, formed by
stoichiometric samples, to the formation of (100)/(001) growth twins
and, finally, even to (001)/(111) growth twins at large $x$ (see below
and fig.\ref{CuRes}). The same twin structures have also been observed
in 123 thin films. From the degree of their structural perfection and
from the number of crystal nuclei of a growth experiment we infer that
the $R$/Ba-ratio of the growing crystal is closely related to the
supersaturation during growth.

The local Pr/Ba-ratio of a given sample with fixed overall $x$ can be
influenced by an additional heat treatment, but the composition of the
resulting crystal is inhomogeneous. For small deviations from the
equilibrium surface a crystal may decompose either into a member of the
$R123ss$ with a different cation ratio, and other stable phases like
BaCuO$_2$, or into two $R$123ss phases with different $R$/Ba ratios.
Larger deviations from equilibrium may lead to a spinodal decomposition
where, for example, the Pr/Ba-ratio varies lateraly and changes with
time. The final state of these decomposition reactions depends on the
initial $R$/Ba-ratio and whether or not this cation ratio is located
inside the four-dimensional volume of the $R123ss$. In the case of a
spinodal decomposition the regions of different Pr/Ba ratios within a
sample are very close to each other (10 - 100 nm), and any integral
chemical analysis will reproduce the composition of the untreated
sample. The regions with a lower Pr/Ba ratio may, however, be large
enough to be superconducting. 

Fig.\ref{CuRes} shows the field-sweep spin echo Cu(1)-spectra of three
crystals. 
The external field is applied along the $a$- and $b$-axes
of the twinned samples. 
In the present case of a large Zeeman splitting the position of the
central line is in first order perturbation theory given by $B_{hf}$
alone, independend of the EFG, and the distance to the satellites
corresponds to the component of the EFG-tensor in field direction. The
spectra are centered at the Cu-resonance of the metal, showing that the
Cu-chain sites carry no moment. The EFG-tensor of Cu(1) with two, three,
and 
\begin{table}
\label{tab1}
\begin{tabular}{llll}
 & Pr$^-$123 & Pr$^0$123 & Pr$^+$123 \\
\hline
at\%Pr & 2      & 2         & 2         \\
at\%Ba & 37     & 30        & 28        \\
at\%Cu & 61     & 68        & 70        \\
\hline
 & 300 mbar air & 1 bar air & 1 bar O$_2$ \\
\hline
$[^\circ$C$]$& 970$\rightarrow$906& 1000$\rightarrow$939& 1000$\rightarrow$948 \\
$[^\circ$C/h$]$ & -0.4          & -0.5           & -0.35          \\
              & quench        & slow cool      & quench         \\
\hline
       & orthor.   & orthor.   & tetr.  \\
(100)/ & (010)     & (010)     & (111) \\
\end{tabular}
\caption{Preparation parameters and characterisation of the crystals.
The flux composition is given in the first block, followed by the
preparation atmosphere and the temperature program,
with a slow decrease at the indicated rate in the given
interval, and a quench or slow cooling to RT. The morphology and
possible growth twins are indicated in the last block.
}
\end{table}
\noindent
four oxygen nearest neighbors is known from the other
$R123$-compounds\cite{Hei3,lut329}. The fourfold coordination is
characterized by $\mid V_{zz}\mid \approx\mid V_{yy}\mid $ or $\eta=0.9$
(see fig.\ref{CuRes}). It is the only one which we could identify in our
crystals, as is expected for filled chains.

The spectrum of sample Pr$^+$123 shows a larger splitting than for
Pr$^0$123 as well as an inhomogeneous broadening of the satellite lines,
but no significant broadening of the central line. The high field
shoulder of the central transitions at 7.4~T can be assigned to the
(111) growth twins mentioned above. The satellite line broadening shows
immediately the presence of disorder in the charge 
coordination of chain site Cu, 
and the unchanged central line proves the absence of
internal magnetic fields at the Cu(1)-site. The substitution of divalent
Ba by Pr induces the inhomogeneity in the charge distribution of the
chain layer detected in the Cu(1)-EFG. The result confirms, therefore,
the previous observations that Pr-rich crystals with $x>0$ form in
oxygen rich atmosphere\cite{Zitat4}. The widths of the central lines
show, on the other hand, that the magnetic structure of the
Cu(2)-sublattice is unchanged, indicating that no ZR singlet hole states
are doped in the CuO$_2$-planes.

The spectrum of the nonstoichiometric crystal Pr$^-$123 is, in contrast,
best described by two subspectra. One with app. 30\% relative intensity
is the same as for the stoichiometric crystal, accordingly the
environment of 1/3 of the Cu(1)-sites is unchanged. A significant
broadenening of the central transitions together with a severe
broadening of the satellites shows the presence of inhomogeneous
magnetic fields together with charge disorder for the majority of the
Cu(1)-sites. The inhomogeneous width of the central transition
corresponds to internal magnetic fields of $\approx 0.2$~T, in full
agreement with the transferred hyperfine fields from the Cu(2)-moments
at the Cu(1)-site ($\approx 0.1$~T) that are known from studies of the
so-called AF-II structure in Al-doped Y123 \cite{bre350}. The defect
structure in this crystal affects, therefore, the magnetic structure of
the Cu(2)-sublattice, as is expected if ZR singlets are doped in the
CuO$_2$-planes. If we assume that only the eight nearest
Cu(1)-neighbors of a hole contribute to this subspectrum we arrive at a
concentration of $\approx 0.2/$unit cell. It is tempting to assign the
signal in this way to Cu(1)-neighbors of localised Zhang-Rice singlets,
but it should be noted that disorder in the stacking sequence of perfect
antiferromagnetic CuO$_2$-planes may lead to the same inhomogeneous
transferred hyperfine fields.

The above interpretation of the Cu(1)-spectra is supported by the
properties of the Pr-resonance, which provides a very sensitive probe of
the van Vleck susceptibility of the $4f^2$-shell and, therefore, of the
crystalline electric field (CEF) from the neighboring
ions\cite{neh352}
 ({\em not} of the EFG at the Pr-nucleus, the
quadrupole splitting is too small to be resolved). 
Let us assume that the Pr-signal is due to Pr on
regular rare earth sites, as discussed in
refs.\cite{neh352,inpreparation}. Comparison of the Pr-spectra in the
three crystals shows immediately that the shape of the line is
independent of $x$ for Pr/Ba-ratios larger than the stoichiometric value.
In contrast, we have scarcely been able to detect any Pr-resonance in
the sample Pr$^-$123. The first observation shows clearly that the
electronic state of the CuO$_2$-Pr-CuO$_2$-trilayer is the same in both
crystals, so the holes are localised in the same states. The second
result shows that the electronic configuration is different in crystals
prepared under conditions favouring Ba-rich phases, even if they are not
superconducting. Taking into account the magnetic broadening at the
Cu(1)-sites this is strong evidence for the presence of Zhang-Rice
singlets. Such a defect will change the CEF of neighboring Pr-ions, and
most probably it will extinguish the contribution of its neighborhood
to the Pr-signal by inhomogeneous broadening or fast relaxation. The
singlets might be induced by a structural modification like the larger
Pr-O(2)-bond length considered by Zou et al. We think, however, that the
correlation with the preparation conditions supports the second
possibility that Ba$^{2+}$ substitutes Pr on the 
\begin{figure}
\psfig{file=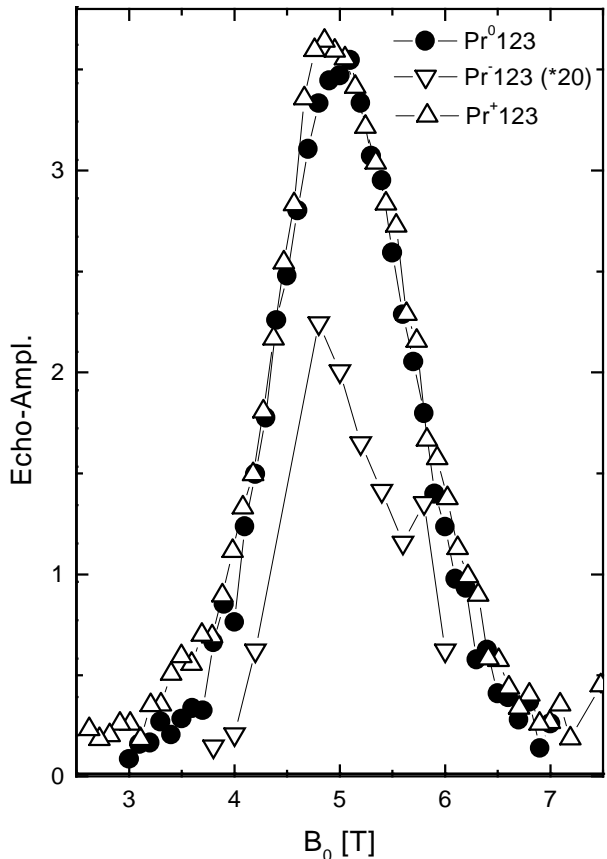, width=8cm, clip=}
\caption{Pr-field-sweep spectra at 102~MHz, 1.3~K, in a field $B_0\|c$. The 
amplitudes are scaled to allow comparison of the line shapes. The signal
in Pr$^-$123 is smaller by more than one order of magnitude.}
\label{PrRes}
\end{figure}
\noindent
RE-site.

We investigated single crystals of the solid solution system
Pr$_{1+x}$Ba$_{2-x}$Cu$_3$O$_7$ grown under preparation conditions
chosen to obtain non-superconducting stoichiometric, Pr-rich, and
Ba-rich specimen. The low temperature NMR-spectra of the Cu(1)- as well
as of the Pr-sites show that the hole states in the CuO$_2$-layers of
the stoichiometric and the Pr-rich crystal are very similar, and we
assign them to the $4f^2-2p\pi$-hybridization band. The spectra of the
crystal prepared under conditions favoring Ba-rich phases show in
contrast a significant magnetic inhomogeneous broadening at the
Cu(1)-site and a vanishingly small Pr-signal, both indications for the
presence of Zhang-Rice singlets. We argue that the occurence of these
singlets at the preparation conditions favoring the Ba-rich phases is a
strong indication that in Pr$_{1+x}$Ba$_{2-x}$Cu$_3$O$_7$ Ba can
substitute Pr under certain conditions. This gives a natural explanation
for the recent observations of intrinsic superconductivity in Pr123
crystals.

{\bf Acknowledgements:} One of us (M.W.P.) was supported during part of
the work by a grant of the Deutsche Forschungsgemeinschaft and
acknowledges the hospitality of the Forschungszentrum
J\"ulich during the preparation of this publication.

\end{document}